\documentclass[journal]{IEEEtran}
\usepackage{epsfig}
\usepackage{graphicx}
\usepackage{graphics}
\usepackage{filecontents}
\usepackage{braket}
\usepackage{cite}
\usepackage{array}
\usepackage{mathtools}
\usepackage{tabularx}
\usepackage{subfigure}
\usepackage{slashbox}

\ifCLASSINFOpdf
\else
\fi
\begin{document}
\title{Multiparty Quantum Key Agreement \\ based on Quantum Secret Direct Communication with GHZ states}
%
%
%

\author{Guo-Jyun~Zeng,
		Kuan-Hung~Chen,
		Zhe-Hua~Chang,
		Yu-Shan~Yang,
		and~Yao-Hsin~Chou*
\thanks{G.-J. Zeng, K.-H. Chen, Z.-H. Chang, Y.-S. Yang, and Y.-H. Chou}
\thanks{Department of Computer Science and Information Engineering}
\thanks{National Chi-Nan University, Taiwan, ROC}
\thanks{G.-J. \;\,Zeng(e-mail: s103321901@ncnu.edu.tw)}
\thanks{K.-H. Chen(e-mail: josh790412@hotmail.com)}
\thanks{Z.-H.\,\,\,Chang(e-mail: s104321520@ncnu.edu.tw)}
\thanks{Y.-S. \:\,Yang(e-mail: s104321519@ncnu.edu.tw)}
\thanks{Y.-H. \:Chou(e-mail: yhchou@ncnu.edu.tw)}
}

\maketitle

\begin{abstract}
	Quantum Key Agreement (QKA) signifies that two or more participants together generate a key and QKA has to satisfy the following conditions: \textbf{1} Every participant can change the key and the key is not decided by any participant individually. \textbf{2} Only participants can know the key; nonparticipants cannot get the key through illegal means. Because of the condition \textbf{1} of participating together, it makes transport inefficient in the current mainstream protocols. They use unicast to exchange messages one by one, so it will considerably limit transmission efficiency and increase cost time spent. This study proposes a protocol based on Multiparty Quantum Secret Direct Communication (MQSDC) with multicast. In addition to satisfying the above conditions, it uses multicast to not only achieve the effect and purpose of QKA, but also to defend against internal and external attacks at the same time. In regard to resource consumption, this study involves linear growth and is more efficient than other mainstream protocols which employ exponential growth.
\end{abstract}

\begin{IEEEkeywords}
	Multiparty Quantum Key Agreement, Quantum Secret Direct Communication, Quantum Key Agreement.
\end{IEEEkeywords}

%
\IEEEpeerreviewmaketitle

\section{Introduction}\label{sec:introduction}
	\IEEEPARstart{Q}{uantum} cryptography has been paid attention since Bennett and Brassard \cite{Bennett1984} proposed the first quantum key distribution (QKD) protocol in 1984. Its security had been proved to be unconditionally secure \cite{Lo1999} \cite{Shor2000}. It is foundation of Quantum cryptography, detect eavesdropper and distribute classical key. The key point is that the security of quantum cryptography is based on quantum theory (such as uncertainty principle and quantum no-cloning theorem \cite{Wootters1982}), rather than the assumption of computation complexity which is the mathematical problems that are hard to be solved (such as discrete logarithm and prime factor decomposition).
	
	Moreover, the quantum theory can be a spear to break the protection of classical cryptography by the parallel computation. The quantum computer can evaluate all the solutions at the same time, and find the exactly one by the superposition principle in quantum theory. The famous algorithms are Deutsch-Jozsa \cite{Deutsch1992} (an algorithm which can distinguish the kind of input function), Shor \cite{Shor1994} (an algorithm which can speed up the prime factor decomposition) and Grover \cite{Grover} (an algorithm which can search the data from unsorted database by $O(\sqrt{N})$). As a result, the classical cryptography will be challenged when the quantum computer is implemented. Therefore, the quantum cryptography is the best way to avoid the destroy from the quantum computer.
	
	The quantum cryptography has been developed over 30 years. It contains four topic which are quantum key distribution (QKD), quantum secret direct communication (QSDC), quantum secret sharing (QSS), and quantum oblivious transfer (QOT). The concept of them evolved from classical cryptography. However, the research of quantum key agreement (QKA) is slow development. The key agreement (KA) is common in classical cryptography, but the QKA was designed by Zhou et al. \cite{Zhou2004} later in 2004. The QKA is a subset from QKD, but the condition of key generator is stricter than QKD. QKA has two important conditions, \textbf{1} Every participant can change the key and the key is not decided by any participant individually, \textbf{2} Only participants can know the key; nonparticipants cannot get the key through illegal means. However, the condition \textbf{1} causes that two-party QKA is hard to extend to multi-party until 2012. Shi and Zhong \cite{Shi2012} proposed the first multi-party QKA (MQKA) by using Bell states. After that, Liu et al. \cite{Liu2012} pointed out the drawback of Shi and Zhong's protocol, and proposed another MQKA protocol by using single states. To 2015, more two-party QKA was proposed, such as \cite{Chong2010,Shen2014,Huang2014,Huang2014a}. And more multi-party QKA was also discussed, such as \cite{Shukla2014,Sun2015a,Sun2015}. Nowadays, the QKA is getting more attention, the researchers is striving to develop and complete.
	
	However, the researchers have a dispute about the condition \textbf{1}, which said every participant can change the key. It means that every participant should join the key generation. Some researchers define the ``join" as ``measurement", such as \cite{Huang2014,Xu2015}. The ``measurement" means that the participants can not change the key by their idea, the key will be determine by random, just like BB84 \cite{Bennett1984}. But, some researchers consider the ``join" to be ``operation", such as \cite{Shi2012,Liu2012,Shukla2014,Sun2015a,Sun2015}. The ``operation" means that the participants can inject their idea of key into the final key. In other word, the second definition is harder to implement than the first. So far, the MQKA protocols \cite{Shi2012,Liu2012,Shukla2014,Sun2015a,Sun2015} are belonged to the ``operation" definition.
	
	Even so, their protocols are ineffective because condition \textbf{1} causes the protocol design to be unicast. All of participants should exchange their operation with the others for final key generator. It is worth mentioning the multicast will better than the unicast. And our protocol is a multicast design which is inspired by multi-party QSDC proposed by Jin et al. \cite{Jin2006} in 2006. It can use multicast to transmit their operations to all participants at once. In this way, our protocol is more efficient than \cite{Shi2012,Liu2012,Shukla2014,Sun2015a,Sun2015}.
	
	This paper is organized as follows. In section \ref{sec:notationDefinition}, the notation definition is defined. Section \ref{sec:theProposedProtocol} is the proposed protocol of this research, first, for easy understand, the two-party QKA (sec. \ref{sec:twoQKA}) will be introduced. And then, multi-party QKA (sec. \ref{sec:MQKA}) are proposed. After that, the key generator (sec. \ref{sec:keyGenerating}) is presented which is the formula for the key without codebook. Section \ref{sec:securityAnalysis} concerns the security analysis of the proposed protocol which are external and internal attack. Section \ref{sec:consumptionComparison} is the consumption comparison. Finally, section \ref{sec:conclusion} concludes this research.

\section{Notation definition}\label{sec:notationDefinition}
	The information carried at quantum computer called quantum bit, abbreviated as qubit. The qubit states are defined as a basis of a 2D plane with the ket notation $\ket{\cdot}$. For example, qubit states $\ket{0}=\left [ 1 \,  0 \right ]^{T}$ and $\ket{1}=\left [  0 \, 1 \right ]^{T}$, it is a standard basis on 2D plane. As a result, the user can define any qubit states according to their requirement. There are two common bases, Z and X basis, $\{ \ket{0}, \ket{1} \}$ and $\{ \ket{+}=\frac{1}{\sqrt{2}}(\ket{0}+\ket{1}), \ket{-}=\frac{1}{\sqrt{2}}(\ket{0}+\ket{1}) \}$.

	The quantum theory uses tensor product $\otimes$ to bind two or more qubits together. And the product can extend the dimension of the qubits, for instance, \\
	$\ket{0}\otimes\ket{0}=\ket{00}=\left [ 1 \cdot \left [ 1 \, 0\right ] 0 \cdot \left [ 1 \, 0 \right ] \right ]^{T}=\left [ 1 \, 0 \, 0 \, 0 \right ]^{T}$.
	
	\subsection{Quantum gate}\label{sec:quantumGate}
		The quantum computer should use the quantum gate to complete the computation. And all of these quantum gates are the unitary operation which is $UU^{*}=I$, where $U^{*}$ is the adjoint of $U$. There are five common single qubit gates (operations) as follows:
		\[
			I=\begin{pmatrix} 1 & 0 \\ 0 & 1 \end{pmatrix},
			X=\begin{pmatrix} 0 & 1 \\ 1 & 0 \end{pmatrix},
			Y=\begin{pmatrix} 0 & -i \\ i & 0 \end{pmatrix},
		\]
		\[
			Z=\begin{pmatrix} 1 & 0 \\ 0 & -1 \end{pmatrix}\;and\;
			H=\frac{1}{\sqrt{2}}\begin{pmatrix} 1 & 1 \\ 1 & -1 \end{pmatrix}.
		\]
		
\begin{figure}[!ht]
			\centering
			\subfigure[]{
				\label{fig:CNOT_a}
				\includegraphics[scale=0.4]{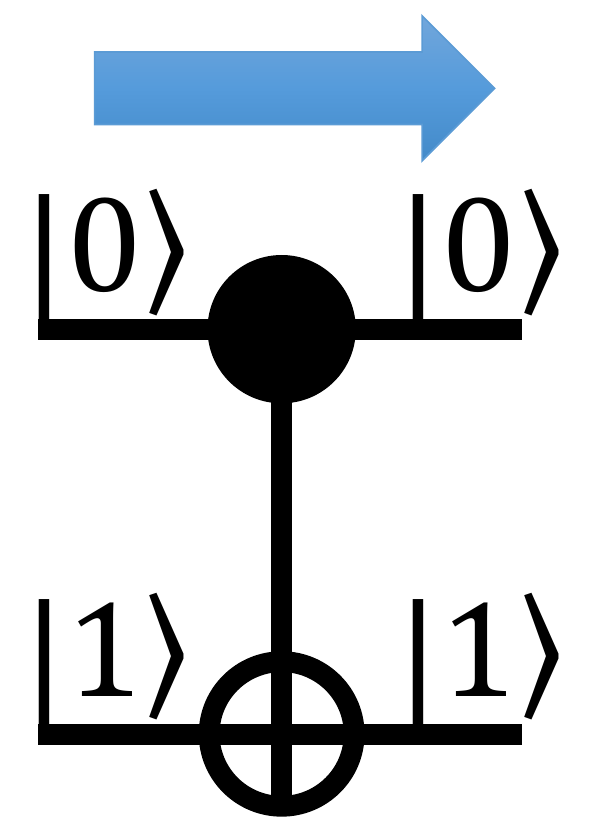}
			}
			\hspace{1cm}
			\subfigure[]{
				\label{fig:CNOT_b}
				\includegraphics[scale=0.4]{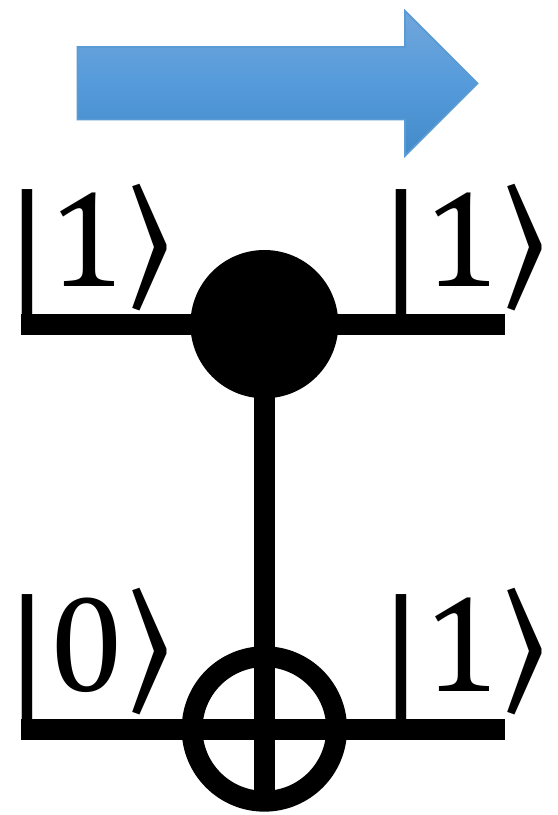}
			}
			\caption{The control-not gate. (a) the control bit doesn't influence the target bit. (b) the target bit performs a not operation by influencing of control bit.}
			\label{fig:CNOT}
		\end{figure}

		Moreover, quantum computer can use a special gate to operate two or more qubits, called control-not gate (\textit{CNOT}, as FIg. \ref{fig:CNOT}). It is composed of control and target bit. The control bit is the input bit which can influence the target bit. And the target bit will change its state according to the signal of control bit. For example, the Fig. \ref{fig:CNOT_a} doesn't influence the target bit when the input bit is $\ket{0}$. On the other hand, the target bit will be perform Not gate, if the control bit is $\ket{1}$, such as Fig. \ref{fig:CNOT_b}. This paper use $CNOT_{A, B}$ to present the control-not gate, where subscript $A$ and $B$ are control and target bit, respectively.
		
	\subsection{Superposition}\label{sec:superposition}
		Different from the classical bit, the qubit contains state $\ket{0}$ and $\ket{1}$ at the same time, such as follow:
		\begin{equation} \label{eq:superposition}
			\ket{\psi}=\alpha\ket{0}+\beta\ket{1}
		\end{equation}
		where the $\left |\alpha \right |^{2}$ and $\left |\beta \right |^{2}$ are the probability to get the state $\ket{0}$ and $\ket{1}$, respectively.
		
	\subsection{Entanglement}\label{sec:entanglement}
		The entanglement is another powerful property of quantum machine. It happens at two or more qubits. In fact, the entangled state is the basis at high dimension. The common two qubits entangled states are called Bell state. It contains four entangles states as follows:
		\begin{equation} \label{eq:bellStates}
			\begin{aligned}
				\ket{\Phi^{\pm}}_{AB}=\frac{1}{\sqrt{2}}(\ket{00}\pm\ket{11})_{AB},\\
				\ket{\Psi^{\pm}}_{AB}=\frac{1}{\sqrt{2}}(\ket{01}\pm\ket{10})_{AB},
			\end{aligned}
		\end{equation}
		where the $A$ and $B$ are the number of the first and second qubit, respectively.
		
		And the three or more entangled states are called GHZ state. For instance, three entangled GHZ states contain 8 states as follows:
		\begin{equation} \label{eq:ghzStates}
			\begin{aligned}
				\ket{\Psi_{000,100}}_{ABC}=\frac{1}{\sqrt{2}}(\ket{000}\pm\ket{111})_{ABC},\\
				\ket{\Psi_{001,101}}_{ABC}=\frac{1}{\sqrt{2}}(\ket{001}\pm\ket{110})_{ABC},\\
				\ket{\Psi_{010,110}}_{ABC}=\frac{1}{\sqrt{2}}(\ket{010}\pm\ket{101})_{ABC},\\
				\ket{\Psi_{011,111}}_{ABC}=\frac{1}{\sqrt{2}}(\ket{011}\pm\ket{100})_{ABC},
			\end{aligned}
		\end{equation}
		where the subscript of $\ket{\Psi}$ is GHZ state number, for example, $\ket{\Psi_{000}}$ present $\frac{1}{\sqrt{2}}(\ket{000}+\ket{111})_{ABC}$.
		
\begin{figure}[!ht]
			\centering
			\subfigure[]{
				\label{fig:BM}
				\includegraphics[scale=0.4]{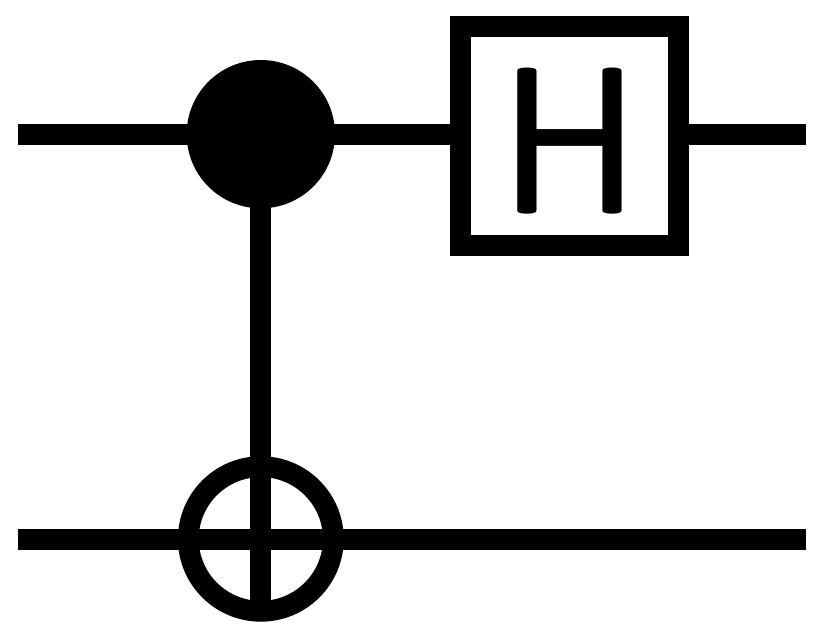}
			}
			\subfigure[]{
				\label{fig:GHZM}
				\includegraphics[scale=0.4]{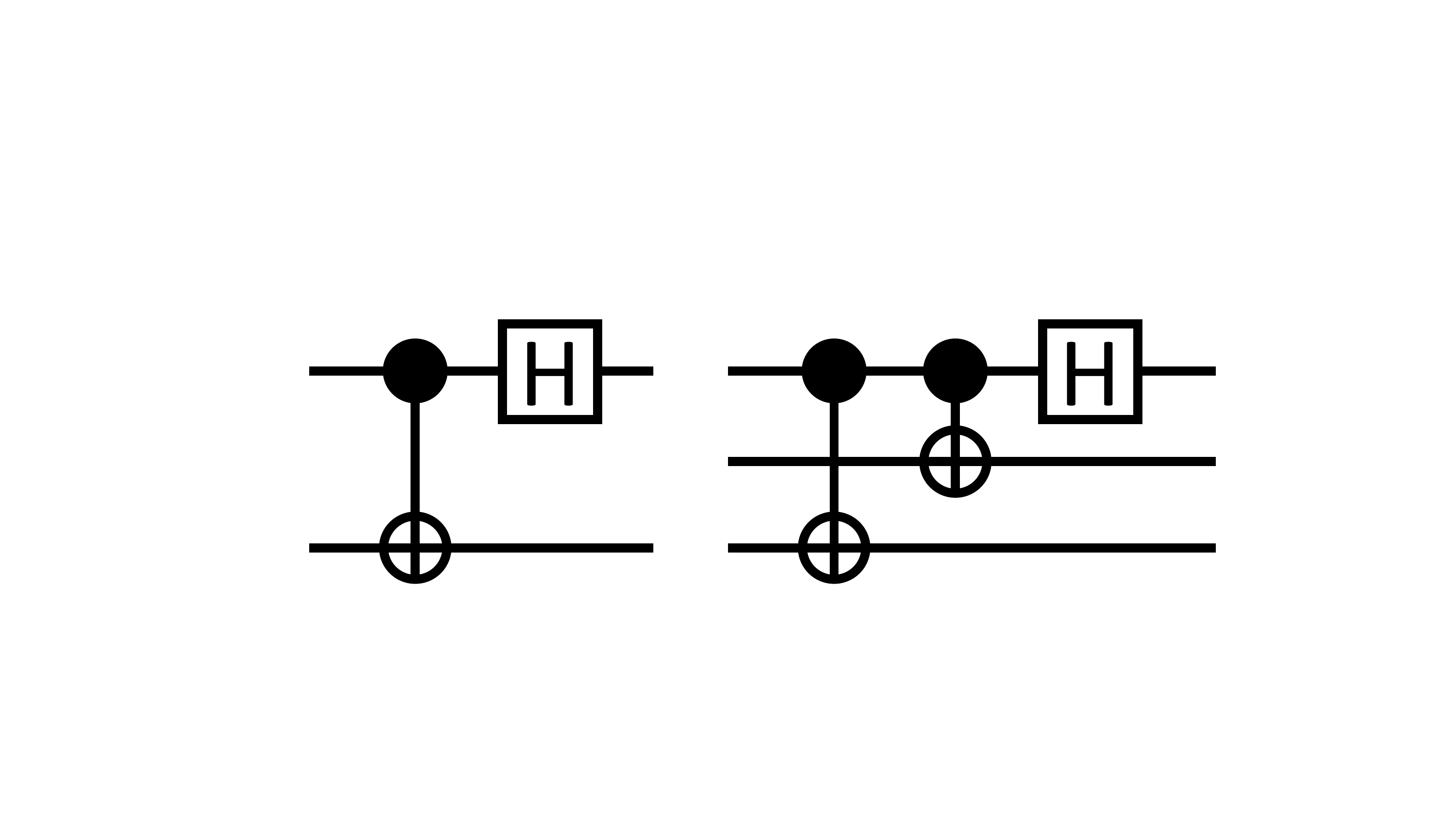}
			}
			\caption{Entangled measurement. (a) Bell measurement. (b) GHZ measurement.}
			\label{fig:entangledMeasurement}
		\end{figure}

		The entangled states should be distinguish by Bell and GHZ measurement. As Fig. \ref{fig:entangledMeasurement}, Fig. \ref{fig:BM} is the Bell measurement. The Bell states should be convert to four states as follows:
		\[
			\ket{\Phi^{+}}\Rightarrow \ket{00},
			\ket{\Phi^{-}}\Rightarrow \ket{10},
			\ket{\Psi^{+}}\Rightarrow \ket{01},
			\ket{\Psi^{-}}\Rightarrow \ket{11}.
		\]
		
		The GHZ measurement is the same as the Bell measurement by using more control-not gate as Fig. \ref{fig:GHZM}, and converts the 8 states to other 8 states as follows:
		\[
			\begin{matrix}
				\ket{\Psi_{000}}\Rightarrow \ket{000},
				\ket{\Psi_{001}}\Rightarrow \ket{001},
				\ket{\Psi_{010}}\Rightarrow \ket{010},
				\ket{\Psi_{011}}\Rightarrow \ket{011},\\
				\ket{\Psi_{100}}\Rightarrow \ket{100},
				\ket{\Psi_{101}}\Rightarrow \ket{101},
				\ket{\Psi_{110}}\Rightarrow \ket{110},
				\ket{\Psi_{111}}\Rightarrow \ket{111}.
			\end{matrix}
		\]
		
		More entangled qubits measurement is called as GHZ measurement too. And it can be wrote as $CNOT_{A, N}CNOT_{A, N-1}...CNOT_{A, B}H$.
		
\section{The proposed protocol}\label{sec:theProposedProtocol}
	This section will introduce our idea sequentially. The first is the easiest case, two-party QKA, and it can also be known as bidirectional QSDC which two participants exchange their secret message. The second is generalizing our protocol to any number of participant. Finally, this protocol proposes a key generator method to quickly exact the key without codebook.
	\subsection{The proposed two-party QKA protocol (basic idea)}\label{sec:twoQKA}
		Our protocol is the improvement from multi-party QSDC protocol by Jin et al. \cite{Jin2006}. In two-party, there are two participants, Alice and Bob, who want to exchange their idea of the secret key. In other word, Alice and Bob has the secret key $Key_{A}$ and $Key_{B}$ respectively. And the final key is $Key_{A}\oplus Key_{B}$. That is, it belongs to condition \textbf{1} that each participant can change the key by their idea. The protocol consists of 5 steps as follows:
		\subsubsection{Step 1 (resource distribution)}\label{sec:BI_step1}
\begin{figure}[!t]
			\centering
			\includegraphics[scale=0.4]{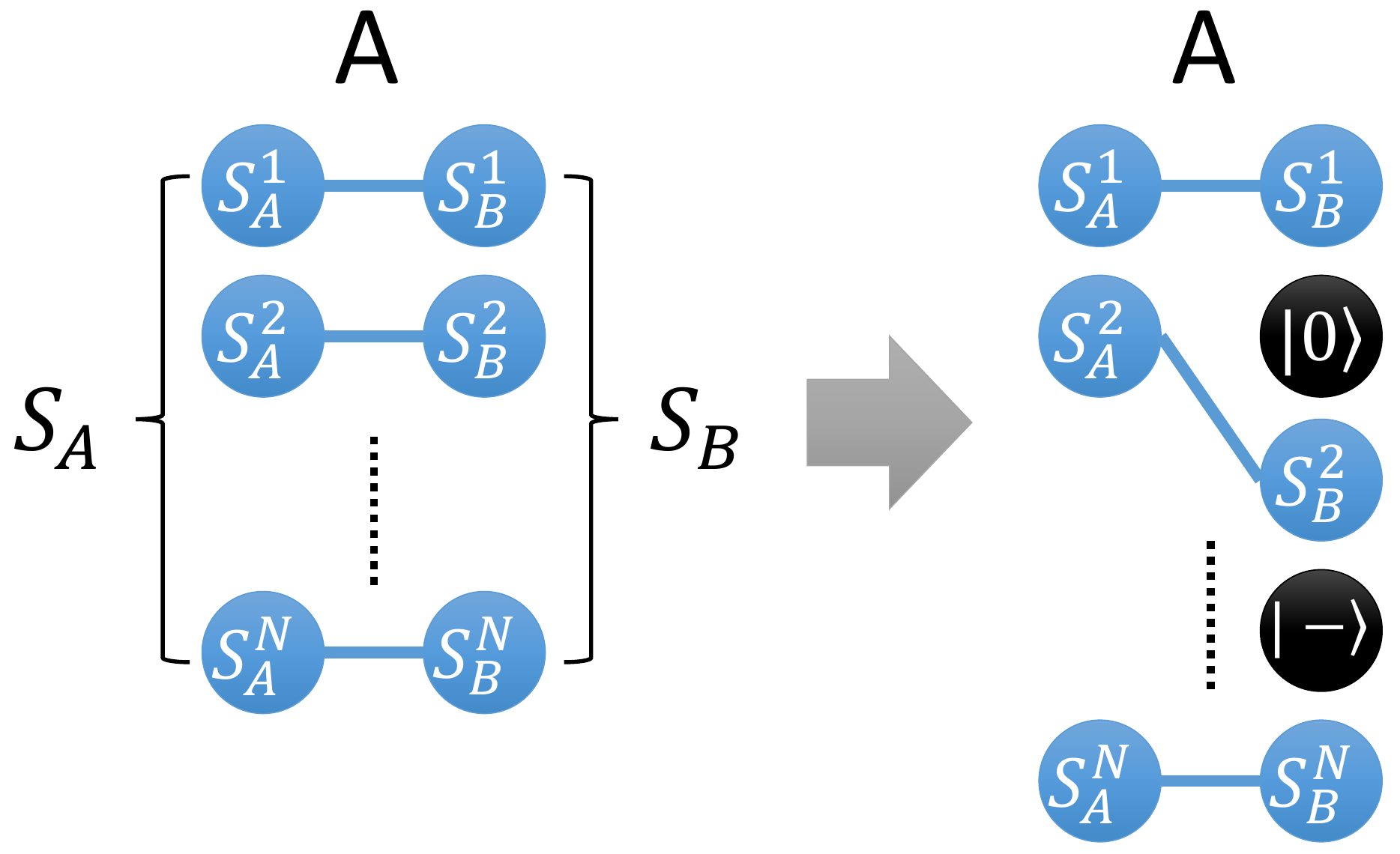}
			\caption{Step 1 of basic idea.}
			\label{fig:BI_S1}
		\end{figure}

		Alice prepares the Bell states sequence, which the each Bell state is $\ket{\Phi^{+}}$. And she split it into two sequences called $S_{A}$ and $S_{B}$ respectively. After that, Alice inserts the single qubits with Z and X-basis into $S_{B}$ for the channel checking, which each qubit is one of two bases. And then, she send $S_{B}$ to Bob. The Fig. \ref{fig:BI_S1} show the whole system state from left to right.
		
		\subsubsection{Step 2 (channel checking)}\label{sec:BI_step2}
		When Bob received the $S_{B}$, Alice tells the position and state of the qubits for the channel checking to him. Then Bob measures these qubits with the bases that is same as Alice prepared. They check these single qubit states. If the error rate is higher than the threshold, this communication should be aborted. Otherwise, they go to the next step. The scenario of this step is showed at left and middle of Fig. \ref{fig:BI_S2_3}.
		
		\subsubsection{Step 3 (self key encryption)}\label{sec:BI_step3}
\begin{figure}[!t]
			\centering
			\includegraphics[scale=0.3]{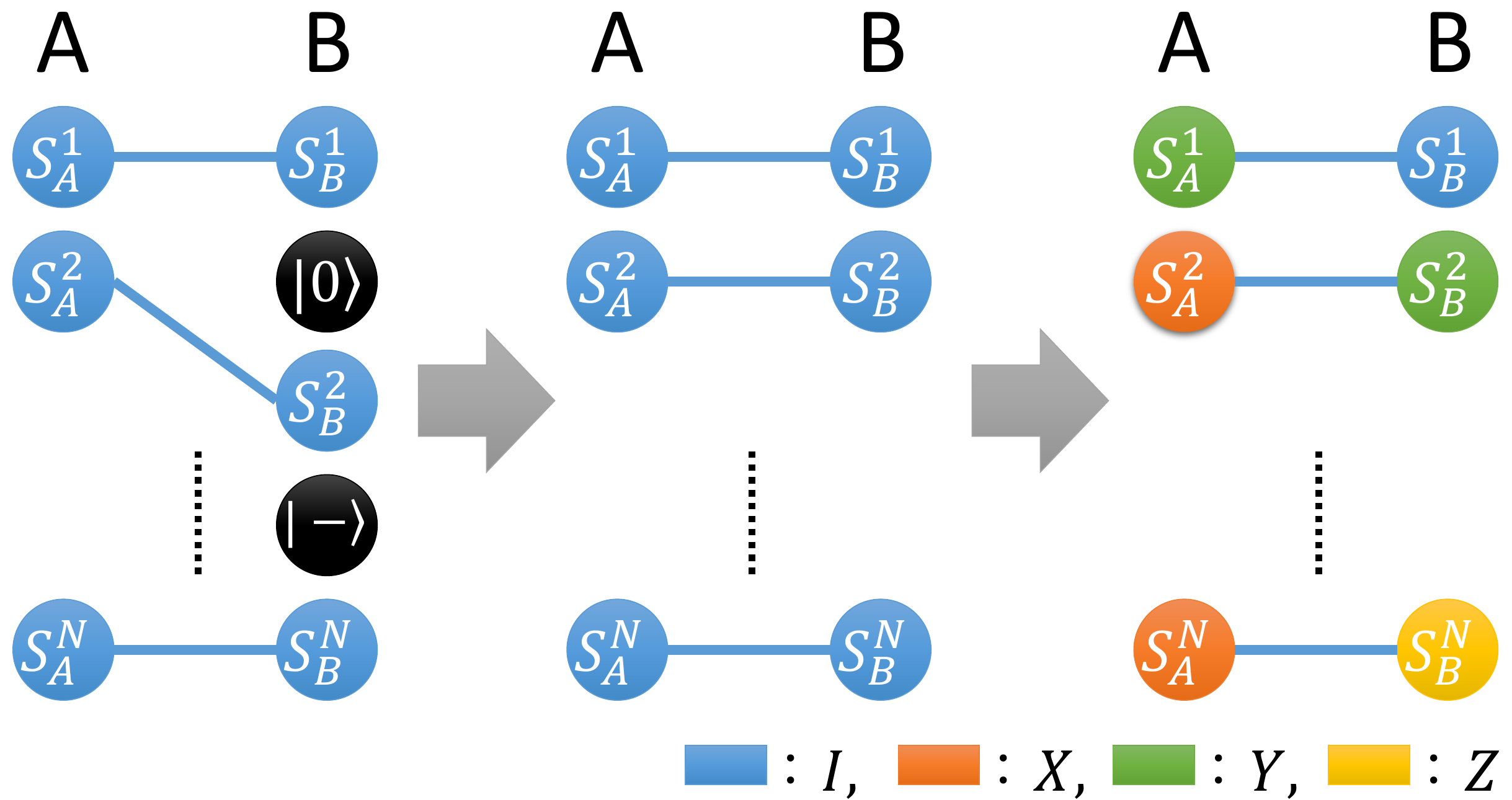}
			\caption{Step 2 and 3 of basic idea.}
			\label{fig:BI_S2_3}
		\end{figure}

		In this step, two roles should be defined first, called leader (L) and follower (F). Leader measures the entangled state and publishes the measurement results to the all followers. In two-party case, Alice is a leader at the number of entangled pair is odd, otherwise, she is a follower. The definition is same as Bob, but even. The leader can performs one of four operations \{$I$, $X$, $Y$, $Z$\} at the qubit hold by him. And the follower can perform one of two operations \{$I$, $X$\} at the qubit hold by him. Leader and follower performs their operations called $M$ according to their self key ``0" and ``1", respectively. The operations $I$, $X$, $Y$ and $Z$ present message ``0", ``0", ``1" and ``1" for leader in two-party case, respectively. And operations $I$ and $X$ presents ``0" and ``1" for follower in two-party case, respectively. It is showed at right of Fig. \ref{fig:BI_S2_3}
		
		\subsubsection{Step 4 (channel checking)}\label{sec:BI_step4}
\begin{figure}[!t]
			\centering
			\includegraphics[scale=0.5]{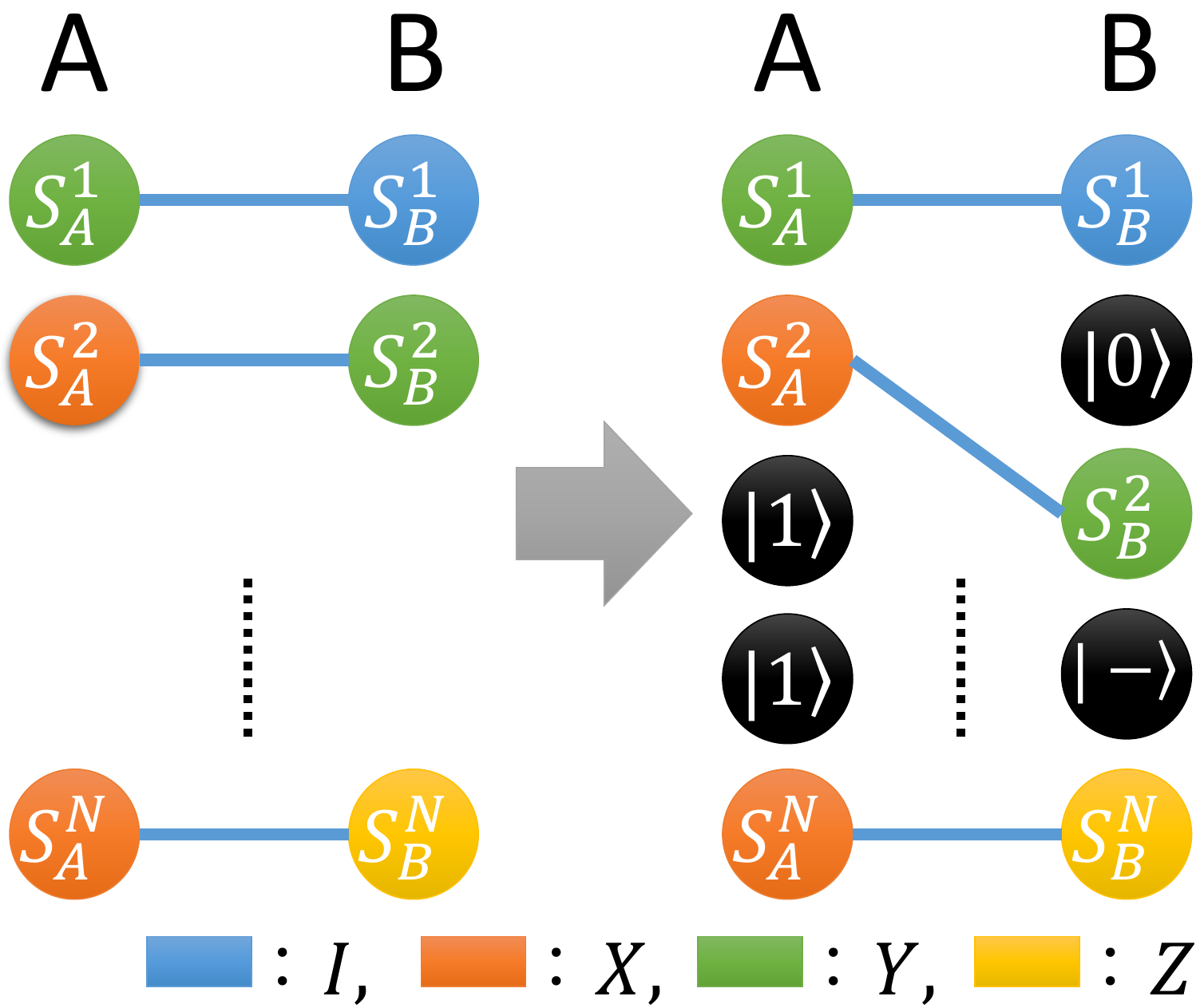}
			\caption{Step 4 of basic idea.}
			\label{fig:BI_S4}
		\end{figure}

		Alice and Bob inserts the qubits into the sequence which will be transferred to another with Z and X-basis, such as Fig. \ref{fig:BI_S4}. And then, Alice sends the sequence to Bob which all the number of qubits are even. And Bob does the same thing with odd. After these two sequences are received by them, they tell the basis and position of these qubits, and perform the measurement on them. If the channel is safe, they go to the next step. Otherwise, they abort this communication.
		
		\subsubsection{Step 5 (secret key generating)}\label{sec:BI_step5}
\begin{figure}[!t]
			\centering
			\includegraphics[scale=0.35]{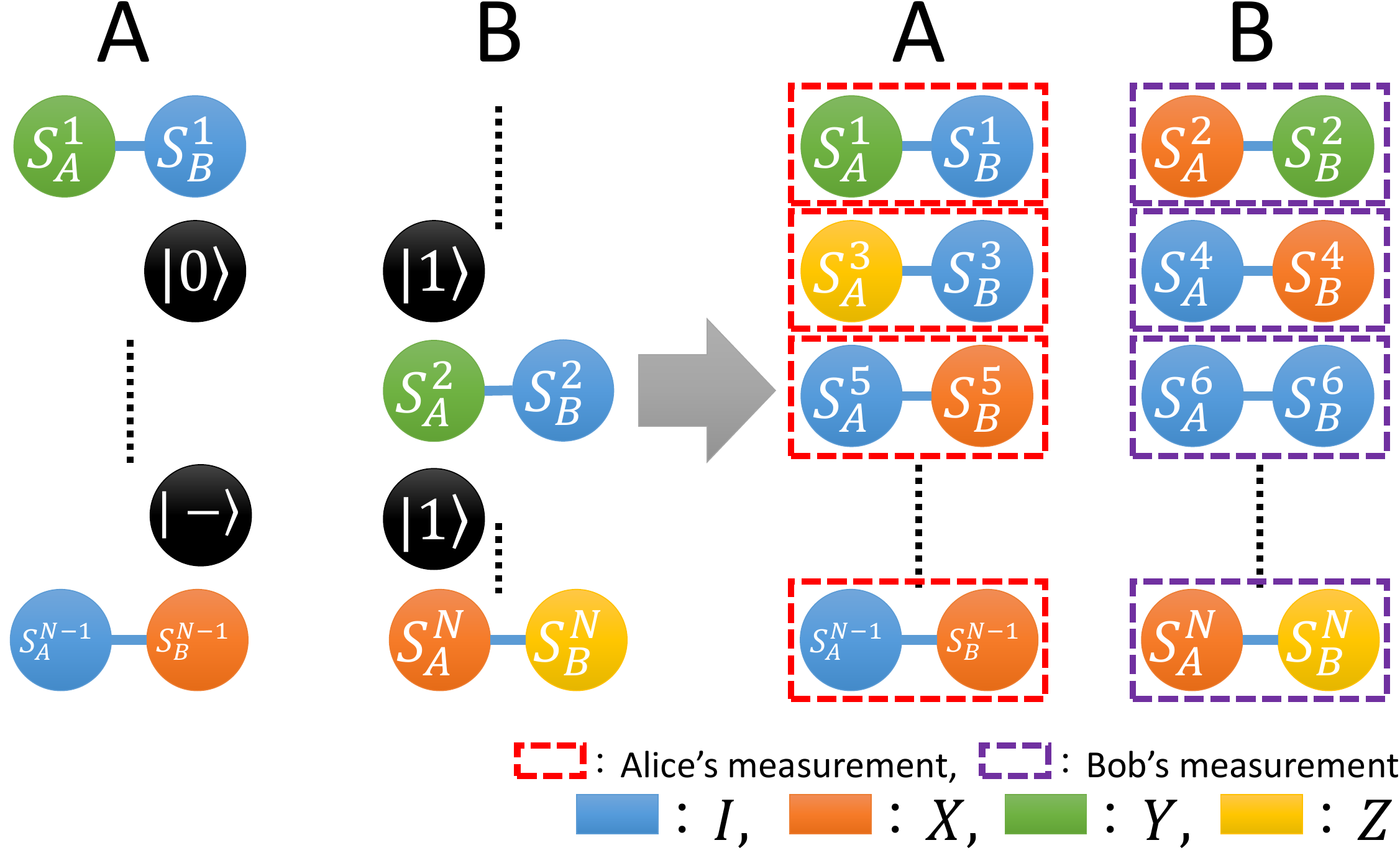}
			\caption{Step 5 of basic idea.}
			\label{fig:BI_S4_5}
		\end{figure}

		Leader performs Bell measurement on the entangled qubits hold by him, and publishes the measurement results to follower, showed in Fig. \ref{fig:BI_S4_5}. Finally, they can exact the operations of another participant did, and decided the final key. For the simple example as Table \ref{tbl:BI_keyGenerating} at a entangled pair, if the Bell measurement result published by leader is $\ket{\Psi^{+}}$, and the operation of follower is $X$, the final key is ``1". According to the measurement result, leader can also know the key is ``1" by his operation $I$. The people who is not a participant can not know the final key, because there are two keys ``0" and ``1" according to the $\ket{\Psi^{+}}$ of Table \ref{tbl:BI_keyGenerating}.
		
		\begin{table}[hbtp]
			\centering
			\tabcolsep=5pt
			\caption{Key generating of two-party QKA}
			\label{tbl:BI_keyGenerating}
			\begin{tabular}{|c|c|c|c|c|}
			\hline
			\backslashbox{F's op.}{L's op.}& $L_{I}$ & $L_{X}$ & $L_{Y}$ & $L_{Z}$ \\
			\hline
			$F_{I}$ & $\ket{\Phi^{+}}=0$ & $\ket{\Psi^{+}}=0$ & $\ket{\Psi^{-}}=1$ & $\ket{\Phi^{-}}=1$ \\ \hline
			$F_{X}$ & $\ket{\Psi^{+}}=1$ & $\ket{\Phi^{+}}=1$ & $\ket{\Phi^{-}}=0$ & $\ket{\Psi^{-}}=0$ \\ \hline
			\end{tabular}
		\end{table}
		
	\subsection{The proposed multi-party QKA protocol}\label{sec:MQKA}
	According to the basic idea of two-party QKA protocol, it can be generalized to multi-party case by defining the relationship between leader and followers. The multi-party case contains 5 steps as follows:
	\subsubsection{Step 1 (resource distribution)}\label{sec:MQKA_S1}
	The participants Alice, Bob, Charlie, ..., Nick will agree a session key. First, Alice prepares the GHZ sequence with the GHZ state as
	\begin{equation*}
		\frac{1}{\sqrt{2}}(\ket{000...0}+\ket{111...1})_{ABC...N},
	\end{equation*}
	where A, B, C, ..., N are presented participant Alice, Bob, Charlie, ..., Nick, respectively. And she splits it to N sequences called $S_{A}$, $S_{B}$, $S_{C}$, ..., $S_{N}$, respectively. Then she inserts the qubits for channel checking with random one of four states \{$\ket{0}$, $\ket{1}$, $\ket{+}$, $\ket{-}$\} for each qubit to $S_{B}$, $S_{C}$, ..., $S_{N}$, respectively. After that, she sends $S_{B}$, $S_{C}$, ..., $S_{N}$ to the Bob, Charlie, ..., Nick, respectively.
	
	\subsubsection{Step 2 (channel checking)}\label{sec:MQKA_S2}
	Alice publishes the basis and position which she prepared for channel checking when other participants received their sequences. All participants publish their measurement result to Alice. They abort this communication, if the error rate is higher than threshold. Otherwise, they go to next step.
	
	\subsubsection{Step 3 (self key encryption)}\label{sec:MQKA_S3}
	Every participants should be a leader alternately. And others should be followers. The rule of operations of leader is (\ref{eq:keyGeneratingRule}). And the follower can only perform one of two operations $I$ and $X$ to present his self ``0" and ``1", respectively. The reason of different number of operations from leader and follower is that a participant should perform Y and Z to extend the entangled states to maximal. For an example of the leader decision, the string of leader is ``ABC...NABC...NABC...". Alice is a leader at $S^{1}_{A}$, $S^{1}_{B}$, $S^{1}_{C}$, ..., $S^{1}_{N}$, and Bob is a leader at $S^{2}_{A}$, $S^{2}_{B}$, $S^{2}_{C}$, ..., $S^{2}_{N}$ and so on, where subscript is presented as number of GHZ states. When all participants have already been leader, it turn to Alice, Bob, Charlie and so on.
	
	\begin{equation} \label{eq:keyGeneratingRule}
		\left\{
			\begin{array}{l}
				\textit{I and Y present ``0",}\\
				\textit{X and Z present ``1", if N is odd}\\ 
				\textit{I and X present ``0",}\\
				\textit{Y and Z present ``1", if N is even}
			\end{array}
		\right.
	\end{equation}
	
	\subsubsection{Step 4 (channel checking)}\label{sec:MQKA_S4}
	After the self key encryption, all participants insert the qubits for channel checking as Alice did at step 1. They send their qubit to the leader under the number of sequence. When all leaders received the qubits from followers, they publish the qubit state and position for the channel checking, and check the error rate. They abort this communication, if the error rate is higher than threshold. Otherwise, they go to next step.
	
	\subsubsection{Step 5 (secret key generating)}\label{sec:MQKA_S5}
	All leaders perform GHZ measurements and publish the measurement outcomes. After the measurement results, all of participants can distinguish the operations of each follower did, and exact the same key similar as Table \ref{tbl:BI_keyGenerating} to complete the agreement. The rule of key exaction is discussed at next section.
	
	\subsection{Key generating}\label{sec:keyGenerating}
	The final key is determined by XOR result of self key of all participants. In this section, two viewpoints will be discussed, which are leader and follower. Under these two viewpoints, this section gives a rule to exact the operation of all participants.
	\subsubsection{viewpoint of leader}\label{sec:keyGeneratingLeader}
		Leader performs GHZ measurement and publishes the measurement outcomes to all followers. He performs not gate on the qubits except a qubit that he performed one of four operations, if his operation is $I$ and $Y$. Then the result $\ket{0}$ represents that the participant performed $I$,and $\ket{1}$ is $X$.
		
		For example, there are three participants Alice, Bob and Charlie. Alice is a leader in this round. If she publishes the measurement outcome is $\ket{\Psi_{110}}_{ABC}=\ket{110}_{ABC}$, and her operation is $Y$, then she performs not gate on the result of qubit B and C, and she gets $\ket{01}_{BC}$. The result $\ket{0}$ and $\ket{1}$ represent operation $I$ and $X$, respectively. According to (\ref{eq:keyGeneratingRule}), she can build final key as $0_{A}\otimes 0_{B}\otimes 1_{C}=1$.
	
	\subsubsection{viewpoint of follower}\label{sec:keyGeneratingFollower}
		When the followers received the measurement outcomes, they observe the qubit result of leader. The leader must perform $Y$ and $Z$ to change his result to be $\ket{1}$. Furthermore, followers perform not gate on their results, if their qubit state of measurement result is different from their operation which is self key. After that, they can distinguish leader's operation which the result is $\ket{0}$ when the operation is $Y$. Otherwise, his operation is $Z$.
		
		Following the example above, the followers are Bob and Charlie, and the GHZ measurement result is $\ket{110}_{ABC}$. In this case, Bob performed $I$ gate, but the result of qubit B is $\ket{1}$. Then he performs $X$ gate to these three qubits to change the entangled state to $\ket{001}_{ABC}$. And he knows that the operation of leader (Alice) is $Y$, and another follower's (Charlie) is $X$. Furthermore, the result of key agreement is $0_{A}\otimes 0_{B}\otimes 1_{C}=1$ according to (\ref{eq:keyGeneratingRule}).
		
\section{Security analysis}\label{sec:securityAnalysis}
	This section discuss two kind of attacks which are external and internal attack. External attack is the discussion which any non-participants want to get the results of key agreement. And internal attack is that the possibility exist or not that there are any participants can determine the results of key agreement without all participants.
	\subsection{External attack}\label{sec:externalAttack}
	There is an eavesdropper called Eve who wants to exact the secret key. She can try to use three common method to test the quantum system and exact the secret key, which are ``intercept-and-resend", ``control-not" and ``fake-participant". According to the idea before, Eve can get each bit of final key if she gets one operation of any participants. Therefore, this section only discusses a interaction between Eve and one of any participants.
		\subsubsection{Intercept-and-resend attack}\label{sec:interceptAndResend}
		Eve intercepts all qubits from followers to leader. And she measures them to try to get the operations performed by followers. However, she doesn't know what the qubits state and position is, prepared at step \ref{sec:MQKA_S4}. She may change the qubit states prepared by followers, and she will be discovered according to the $\frac{1}{4}$ probability with single qubit \cite{Bennett1984}.
		
		\subsubsection{Control-not attack}\label{sec:ControlNot}
		Eve can try multiple control-not gates on transmission qubits proposed by Gao et al. \cite{Gao2010}. She can know that odd or even operation $X$ performed from followers, if she can know the qubit position of self key transmission. For example, three-party QKA protocol, Alice is a leader in this round with inital state $\ket{\Psi_{000}}_{ABC}$, and Bob and Charlie performs $X$ and $I$ on qubit B and C, it changes the entangled state into $\ket{\Psi_{010}}_{ABC}$ respectively. Then they resend qubit B and C to Alice. During the transmission period, Eve steals them and performs $CNOT_{B, E}$ and $CNOT_{C, E}$ with her single qubit $\ket{0}_{E}$, if she can filter out the qubits for channel checking. She can get $\ket{1}$ which means qubit B $\oplus$ C.
		
		However, she can not know which qubits are for channel checking, and she may influence these qubits and then be detected. Following above, for example, Charlie resends qubit with $\ket{+}_{C'}$ for channel checking at the same position, and Bob resends entangled qubit to Alice. Eve steals and entangles her qubit $\ket{0}$ as above, which changes whole quantum system to be (\ref{eq:multipleCNOTAttack}). As a result, qubit C' and E are entangled which means qubit C' may be $\ket{-}$ at channel checking. They can discover Eve, if the measurement result is $\ket{-}$. Furthermore, this way can not detect Eve, if the C' is prepared with Z-basis, because qubit C' doesn't entangle with E. It reduces the detection rate from 100\% to 25\% with a qubit for channel checking.
		
		\begin{equation}\label{eq:multipleCNOTAttack}
			\begin{array}{l}
				\ \ \ \frac{1}{\sqrt{2}}(\ket{010+0}+\ket{101+0})_{ABCC'E}\\
				\ \ \ \ \ =\ \;\;\ \ \frac{1}{2}
				\begin{pmatrix}
					\ \ \;\ket{01000}+\ket{01010}\\
					+\ket{10100}+\ket{10110}
				\end{pmatrix}_{ABCC'E}\\
				\ \overset{CNOT_{B, E}}{\Rightarrow}
				\frac{1}{2}
					\begin{pmatrix}
						\ \ \;\ket{01001}+\ket{01011}\\
						+\ket{10100}+\ket{10110}
					\end{pmatrix}_{ABCC'E}\\
				\,\overset{CNOT_{C', E}}{\Rightarrow}
				\frac{1}{2}
					\begin{pmatrix}
						\ \ \;\ket{01001}+\ket{01010}\\
						+\ket{10100}+\ket{10111}
					\end{pmatrix}_{ABCC'E}\\
				\ \ \ \ \ =\ \ \ \frac{1}{\sqrt{2}}
					\begin{pmatrix}
						\ \ \;\ket{010}\frac{1}{\sqrt{2}}(\ket{01}+\ket{10})\\
						+\ket{101}\frac{1}{\sqrt{2}}(\ket{00}+\ket{11})
					\end{pmatrix}_{ABCC'E}
			\end{array}
		\end{equation}
		
		\subsubsection{Fake-participant attack}\label{sec:fakeParticipant}
		In this kind of attack, Eve camouflages as one of all participants, and she steals the qubit sequence from Alice and sends single or Bell entangled qubit sequence  to the participant as Alice. After the participant resends his encrypted sequence, Eve reads out the key and encrypts self key of the participant. Then she sends them to the leader in the round. After the measurement result is published from leader, all participant can agree final key normally.
		
\begin{figure}[!t]
			\centering
			\includegraphics[scale=0.5]{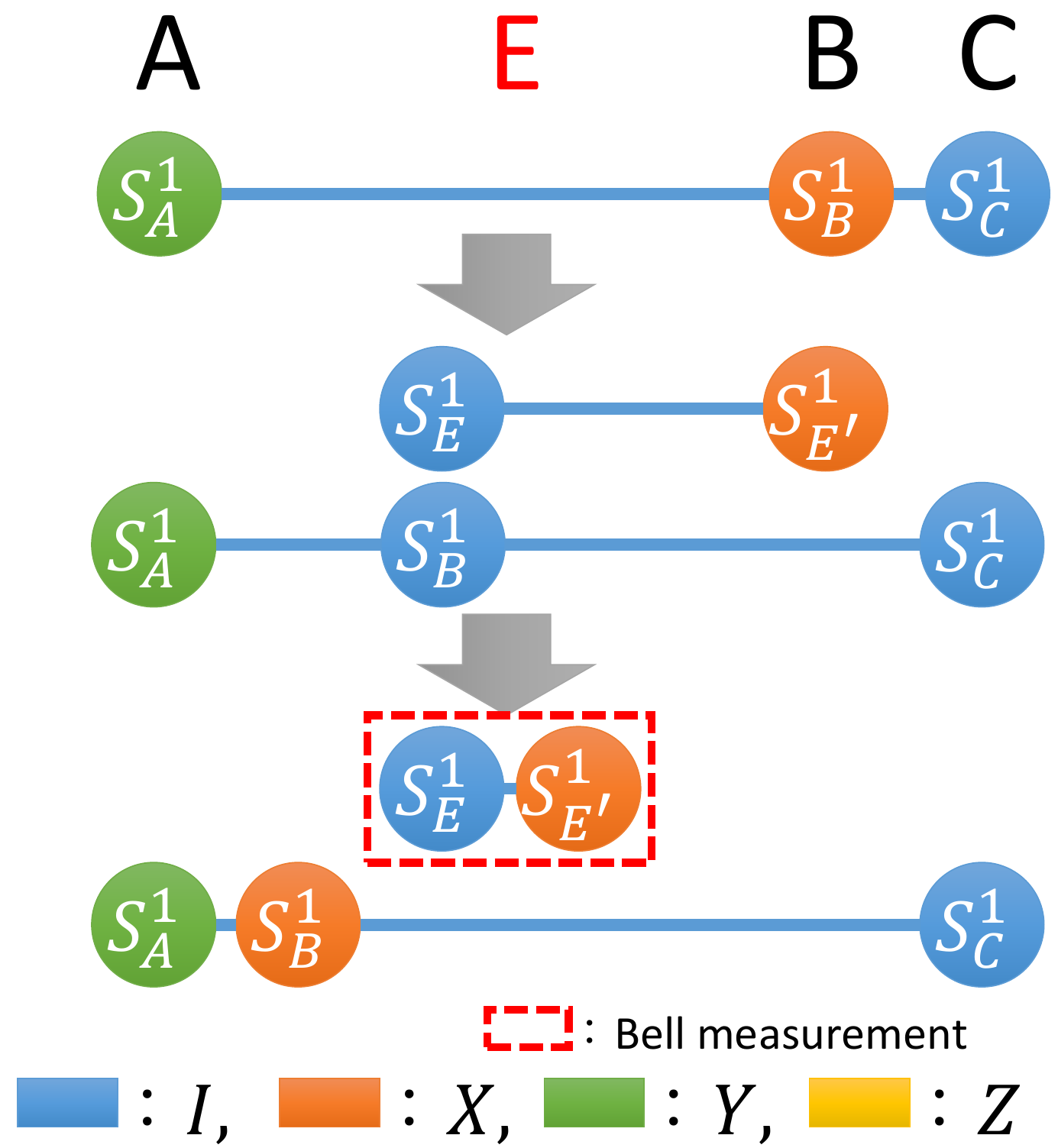}
			\caption{Fake participant attack.}
			\label{fig:fakeParticipant}
		\end{figure}

		For example, three-party case, Alice is a leader in the round, she sends qubit B and C to Bob and Charlie respectively. When the qubit B is transmitted, Eve steals them and sends single or entangled qubit to Bob showed as top of Fig. \ref{fig:fakeParticipant}. Fig. \ref{fig:fakeParticipant} takes entangled state as example. After Bob encrypts his self key at qubit E', he sends it to Alice. Then Eve steals it again, and performs Bell measurement on qubit E and E' to read out Bob's self key showed as middle of Fig. \ref{fig:fakeParticipant}. Finally, Eve encrypts Bob's self key to qubit B and sends it to Alice as normal participant does showed as bottom of Fig. \ref{fig:fakeParticipant}. In this study, Eve can not have enough power to cover all information from any participants. So she can not palm off as a participant completely. Therefore, Eve doesn't know the position of qubits for channel checking. She will be detected at step 4.
		
	\subsection{Internal attack}\label{sec:internalAttack}
	As a result of \ref{sec:theProposedProtocol}, an entangled state can only generate 1 bit key. The key string should be determined by a sequence of entangled states. In our protocol, the leader has capability to determine the key because he can publish the result which he wishes the key. For example, Alice is a leader in two-party case, she can wait the qubit from Bob. After that, she performs Bell measurement first, and reads out the self key of Bob. Then she publish a  measurement outcome according to her idea of final key. For instance, the entangled state of beginning is $\ket{\Phi^{+}_{AB}}$, and Bob performs $X$ gate to present his self key ``1" on qubit B. Then he resends qubit B to Alice. After that, Alice performs Bell measurement and changes the result to be $\ket{\Psi^{+}}$. She publishes $\ket{\Psi^{+}}_{AB}$ and $\Phi^{+}_{AB}$, if she want the final key is ``1". That is, this protocol is designed as leader and followers. All of participants have to be leader alternately. Leader gets chance to determine the final key at a bit. However, he can not determine whole key string.
	
	This protocol discuss a internal attack, ``collusion". In the situation, the subset of participants wants to determine whole key string. Followers can not determine whole key string, even a bit. Since their self key are always encrypted before the leader. Moreover, leader can determine a bit, so the situation is same as that he cooperate with any participants. Therefore it can be showed that \textbf{if any one participant wants to determine whole key string (results of key agreement), he should cooperate with all participants}.

\section{Consumption comparison}\label{sec:consumptionComparison}
	This protocol will compare with 5 current MQKAs by number of ``transmission", ``qubit measurement", ``qubit for channel checking", and ``transmission delay". This study will compare with 5 current MQKA protocols which are ``Shi and Zhong \cite{Shi2012}", ``Liu et al.\cite{Liu2012}", ``Shukla et al. \cite{Shukla2014}", ``Sun et al. 1 \cite{Sun2015}" and ``Sun et al. 2 \cite{Sun2015a}", where Sun et al. proposed two MQKA protocols in same year, so the first is called Sun et al. 1 and the second is called Sun et al. 2, respectively. Sun et al. 2 was the improvement of Shen et al. \cite{Shen2014} in multi-party case. Following these 4 indexes, the computation of consumption will be described as follows:
	\subsection{Number of transmission}\label{sec:numberTrans}
	This subsection discusses number of transmission by 2-bits key agreement. Each qubit is counted the number of transmission from all participant without the qubits for channel checking. The detail of counting is as follows:
	\subsubsection{Shi and Zhong's MQKA protocol \cite{Shi2012}}
		Shi and Zhong's MQKA protocol takes qubit transmission at step 4 as a transmission round. The protocol should takes $N$ rounds for transmission, and each round takes $N$ transmissions, where $N$ is number of participant. As a result, the total transmission number is $N\times N=N^{2}$.
	\subsubsection{Liu et al.'s MQKA protocol \cite{Liu2012}}
		Liu et al.'s MQKA protocol takes qubit transmission at step 3, each participant sends their qubit sequences to others ($N-1$). As a result, the total transmission number is $N\times (N-1)$.
	\subsubsection{Shukla et al.'s MQKA protocol \cite{Shukla2014}}
		Shukla et al.'s MQKA protocol takes qubit transmission at step 2 as a round. They change the operation set in every round. Each participant should send their qubit sequences to others until the qubit sequence travelled all participants. However, the protocol can only agree 1 bit at a key agreement. As a result, the total transmission number is $2\times N\times N$.
	\subsubsection{Sun et al.'s MQKA protocol 1 \cite{Sun2015}}
		Sun et al.'s MQKA protocol 1 takes qubit transmission at step 1 as a round. In every round, all participants send 2 qubit sequences to the previous and following participants transmission respectively. The transmission is finished until each two sequence reach $\left \lceil (N-1)/2 \right \rceil \times 2\times N\times 4$ transmission, where the ceil and 2 is total number of previous and following participant transmission, the last 4 means that each transmission should transmit 4 sequences.
	\subsubsection{Sun et al.'s MQKA protocol 2 \cite{Sun2015a}}
		Sun et al.'s MQKA protocol 2 takes qubit transmission at step 2 as a round. Every round, all participants send their travelling qubit sequence to the following participant. After all participants encrypted their self key, the sequence should be sent to the participant who prepared the sequence. As a result, the total transmission is $N\times N=N^{2}$.
	\subsubsection{Our proposed protocol}
		Our protocol takes step 1 and 4 as a round, respectively. In these two round, the qubit sequences was sent to all participants and resent to leader, respectively. However, each entangled qubits generate 1 bit key. As a result, the total transmission is $(N-1)\times 2\times 2$.
	\begin{figure}[!t]
		\centering
		\includegraphics[scale=0.5]{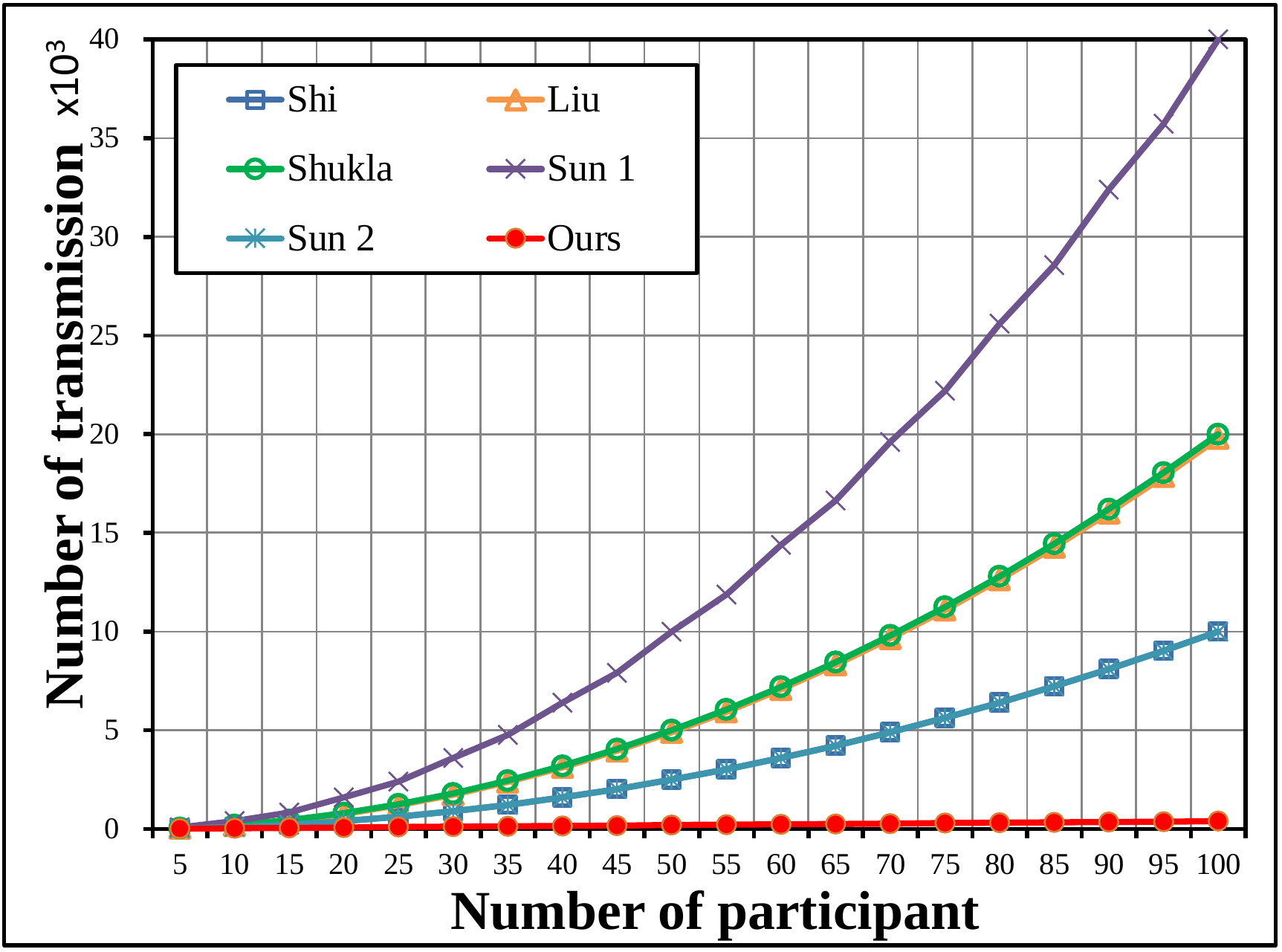}
		\caption{Comparison of number of transmission}
		\label{fig:number_trans}
	\end{figure}
	
	As a result of Fig. \ref{fig:number_trans}, the number of transmission of all current MQKA protocols is $N^{2}$, because of unicast transmission. Therefore, the number of transmissions are higher than this study. Fig. \ref{fig:number_trans} also shows us that multicast transmission is more efficient than unicast.
	
	\subsection{Number of qubit measurement}\label{sec:numberMeasure}
	After 2-bits key agreement, the number of qubit measurement should be discussed. Each participant measures these qubit by specific measurement method such as Bell, GHZ, and cluster measurement. After the measurement, they finish the agreement process. However, the cost is high if the protocol takes more qubit measurement. This section discusses that each protocol takes how many measurement during the key agreement without counting the qubits for channel checking. Assume that 2-bits key is agreed, and each qubit takes a measurement count, the count is shown as follows:
	\subsubsection{Shi and Zhong's MQKA protocol \cite{Shi2012}}
		Shi and Zhong's MQKA protocol only takes a Bell measurement for their key generating at last step of each round. Every participant takes a Bell measurement at a round and tell the measurement result to a participant for each round. Every Bell measurement should be counted 2 in a round. After $N$ rounds, all participants can get the same key. As a result, the total transmission number is $N\times N\times 2$, where the last 2 is that Bell measurement counting.
	\subsubsection{Liu et al.'s MQKA protocol \cite{Liu2012}}
		Liu et al.'s MQKA protocol takes a single qubit measurement by each participant at step 5 as a round. In a round, all participants should measure the qubit sequence sent from $N-1$ participants. However, each round, all participant can agree 1 bit key. As a result, the total transmission number is $(N-1)\times N\times 2$.
	\subsubsection{Shukla et al.'s MQKA protocol \cite{Shukla2014}}
		Shukla et al.'s MQKA protocol takes a Bell measurement by each participant at step 8. Every participants perform Bell measurement on the entangled qubits and generate 1 bit key. As a result, the total transmission number is $2\times N\times 2$, where the first 2 is Bell measurement counting and the last 2 is 2 bit key.
	\subsubsection{Sun et al.'s MQKA protocol 1 \cite{Sun2015}}
		Sun et al.'s MQKA protocol 1 takes 3 Bell measurements by each participant at step 9, and agrees 2 bit key. 3 Bell measurements counts 6 times measurements. And all participants should perform the measurement. As a result, the total transmission number is $3\times N\times 2$.
	\subsubsection{Sun et al.'s MQKA protocol 2 \cite{Sun2015a}}
		Sun et al.'s MQKA protocol 1 takes a cluster measurement by each participant at step 6, and agrees 2 bit key. A cluster measurement counts 4 times measurements. And all participants should perform the measurement. As a result, the total transmission number is $4\times N$.
	\subsubsection{Our proposed protocol}
		Our protocol takes a GHZ measurement by a leader at step 5, and agrees 1 bit key. A GHZ measurement counts $N$ times measurements. As a result, the total transmission number is $N\times 2$.
	
	\begin{figure}[!t]
		\centering
		\includegraphics[scale=0.5]{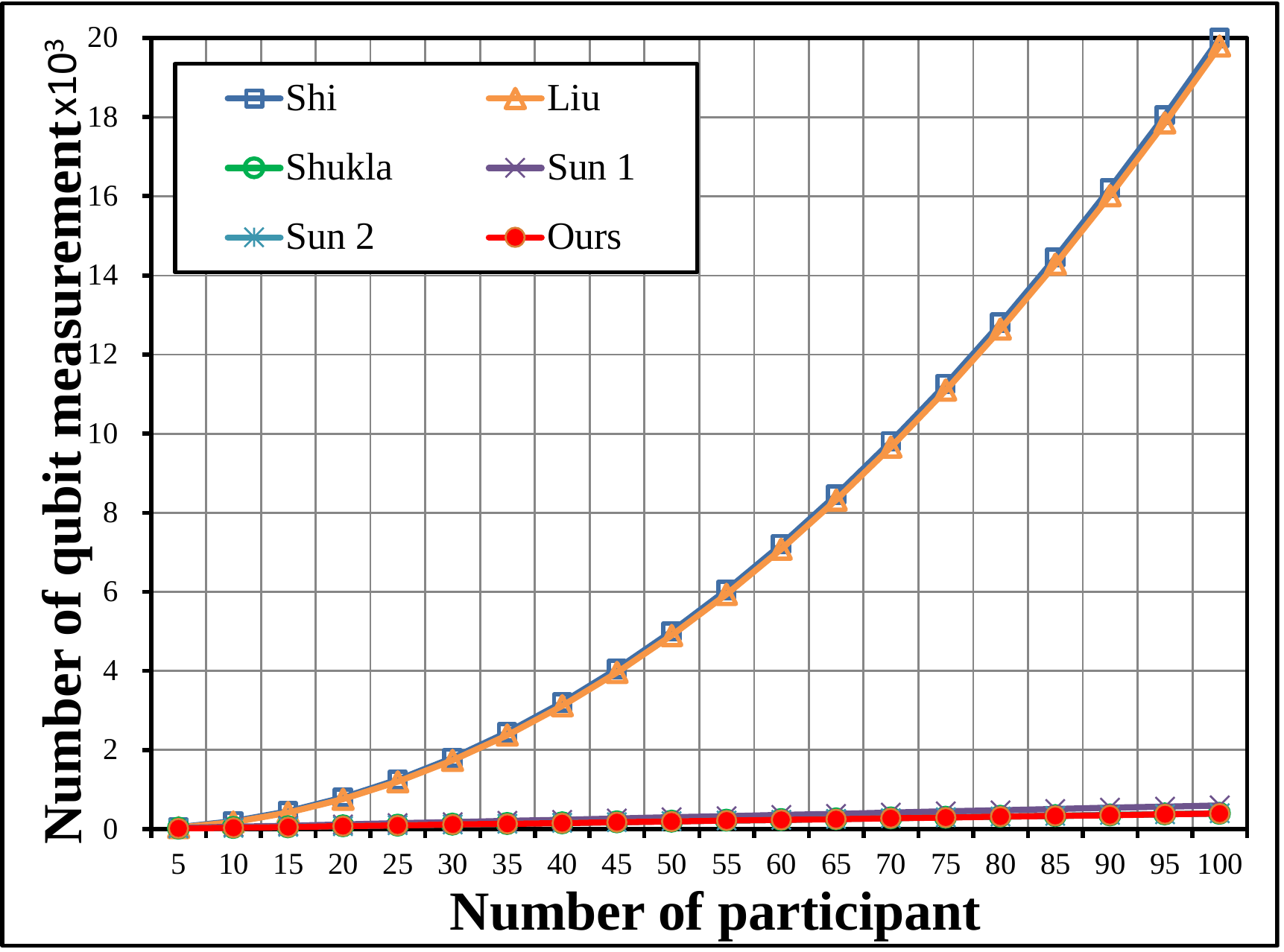}
		\caption{Comparison of number of qubit measurement}
		\label{fig:number_measure}
	\end{figure}
	
\begin{figure}[!t]
		\centering
		\includegraphics[scale=0.5]{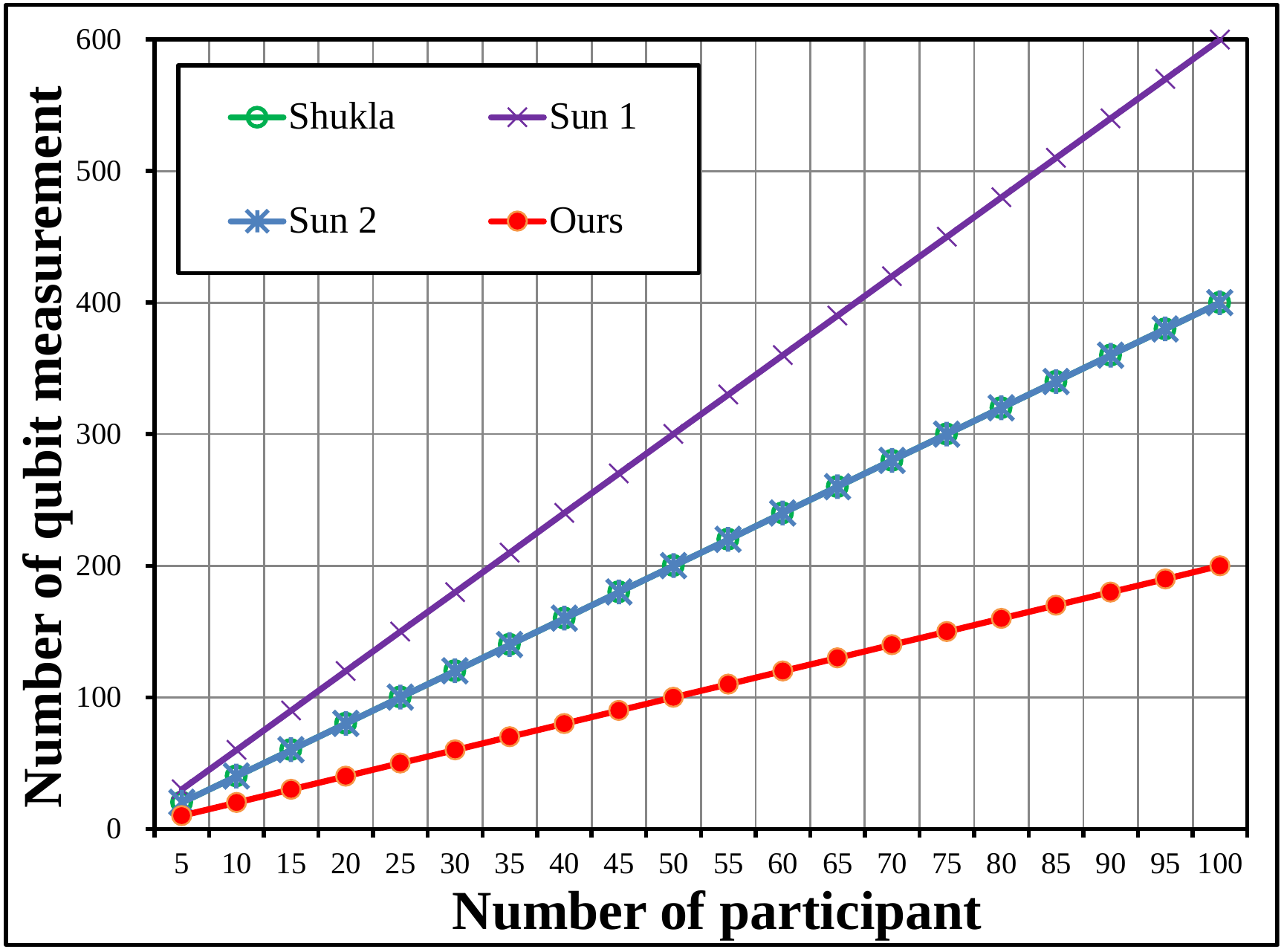}
		\caption{Comparison of number of qubit measurement of \cite{Shukla2014,Sun2015,Sun2015a}}
		\label{fig:number_measure2}
	\end{figure}

	The unicast transmission means that each participant should transmit and measure the qubits. Therefore, the number of qubit measurement is often $N*M$, where $M$ is qubit for key generating. In some protocols, $M$ is constant \cite{Shukla2014,Sun2015,Sun2015a} but some is variable such as \cite{Shi2012,Liu2012}. This study, $M$ is constant, and the performance is a little better than other constant protocols shown in Fig. \ref{fig:number_measure}. However, this study can transmit all entangled qubits in a transmission, and the others should process these qubits round by round. The detail comparison is shown in Fig. \ref{fig:number_measure2}. As a result of comparison of \cite{Shukla2014,Sun2015,Sun2015a},our protocol is better than the others. The advantage of constant $M$ can reduce the transmission delay in our protocol which discuss at \ref{sec:transDelay}.
	
	\subsection{Number of qubit for channel checking}\label{sec:numberChecking}
	The qubit for channel checking should be discussed on the sequences transmission. These sequences are inserted into the qubit for channel checking. In this section, all size of sequences are composed by 100 qubits. There are 10 qubits for channel checking among a sequence. In fairness, each protocol agree 180 length of key (some protocols have to transmit a sequence in a transmission, but some protocols take two sequences or more), the discussion is as follows:
	\subsubsection{Shi and Zhong's MQKA protocol \cite{Shi2012}}
		Every participant of Shi and Zhong's MQKA protocol takes $N$ transmissions. And every transmission is inserted into 10 qubits for channel checking. As a result, the total qubits for channel checking are $N\times N\times 10$.
	\subsubsection{Liu et al.'s MQKA protocol \cite{Liu2012}}
		Every participant of Liu et al.'s MQKA protocol takes $N-1$ transmissions. And every transmission are inserted into 10 qubits for channel checking, where each sequence agrees $100-10$ bit key. Under the requirement of 180 bit key agreement, the protocol should be implemented twice. As a result, the total qubits for channel checking are $(N-1)\times N\times 10\times 2$.
	\subsubsection{Shukla et al.'s MQKA protocol \cite{Shukla2014}}
		Shukla et al.'s MQKA protocol is similar to Shi and Zhong \cite{Shi2012} but only 1 bit key agreement. As a result, the total qubits for channel checking are $N\times N\times 10\times 2$.
	\subsubsection{Sun et al.'s MQKA protocol 1 \cite{Sun2015}}
		Every participant of Sun et al.'s MQKA protocol 1 takes $\left \lceil (N-1)/2 \right \rceil \times 2$ transmissions. And each transmission has to transmit 4 sequences. Every sequence is inserted into 10 qubits for channel checking. As a result, the total qubits for channel checking are $\left \lceil (N-1)/2 \right \rceil \times 2 \times N\times 4\times 10$.
	\subsubsection{Sun et al.'s MQKA protocol 2 \cite{Sun2015a}}
		Every participant of Sun et al.'s MQKA protocol 2 takes $N$ transmissions. And every transmission is inserted into 10 qubits for channel checking. As a result, the total qubits for channel checking are $N\times N\times 10$.
	\subsubsection{Our proposed protocol}
		Our protocol takes $N-1$ transmissions. Every transmission are inserted into 10 qubits for channel checking and agree 1 bit key. Therefore it take 2 times transmissions. As a result, the total qubits for channel checking are $(N-1)\times 2\times 2\times 10$ for 180 bit key agreement.
	
	\begin{figure}[!t]
		\centering
		\includegraphics[scale=0.5]{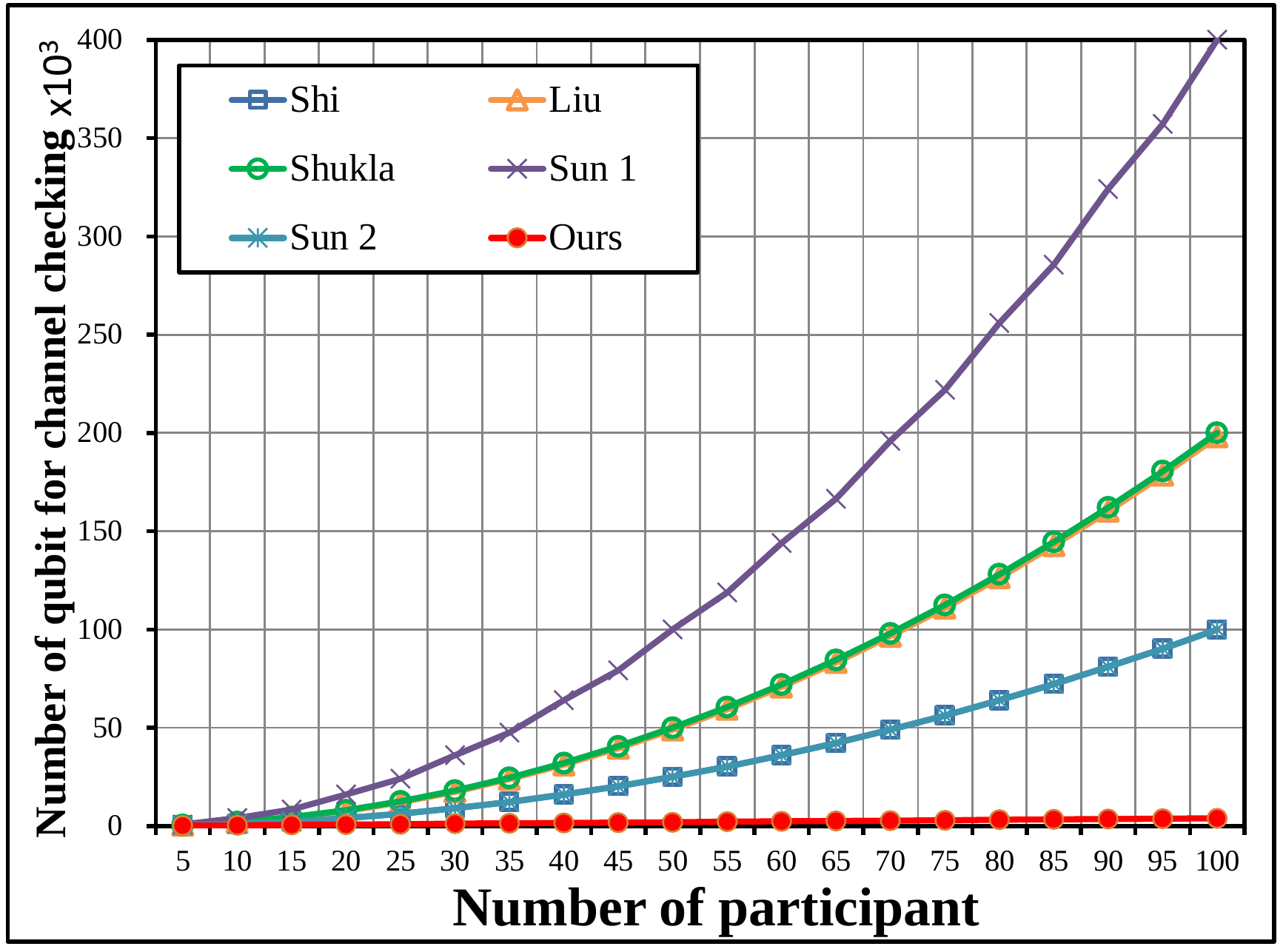}
		\caption{Comparison of number of qubit for channel checking}
		\label{fig:number_check}
	\end{figure}

	\subsection{Transmission delay}\label{sec:transDelay}
	Every transmission should take a time unit for transmitting any rounds and sequences, which the time unit can be nanosecond or millisecond, it is decided by quality of channel. If each round and sequence should be synchronize, the delay is high. This section discusses time delay of each round and sequence transmission for agreeing 2 bit key, the discussion is shown as follows:
	\subsubsection{Shi and Zhong's MQKA protocol \cite{Shi2012}}
		Shi and Zhong's MQKA protocol should take $N$ rounds for helping each participant to agree 2 bit key. After synchronization of each round, all participants can continue to the next 2 bit key. Every round should take a time unit. As a result, the total transmission delay is $N$ time unit.
	\subsubsection{Liu et al.'s MQKA protocol \cite{Liu2012}}
		Liu et al.'s MQKA protocol can transmit all qubit sequences in one round, and each qubit can agree 1 bit key. Under the requirement of  2 bit key, the protocol should take 2 rounds. As a result, the total transmission delay is 2 time unit.
	\subsubsection{Shukla et al.'s MQKA protocol \cite{Shukla2014}}
		The calculation of transmission delay of Shukla et al.'s MQKA protocol is similar to Shi and Zhong \cite{Shi2012}. However, they only agree 1 bit key. As a result, the total transmission delay is $N\times 2$ time unit.
	\subsubsection{Sun et al.'s MQKA protocol 1 \cite{Sun2015}}
		Every participant of Sun et al.'s MQKA protocol 1 should take $\left \lceil (N-1)/2 \right \rceil \times 2$ for the previous and following participants as rounds. All participant have to synchronize every transmission in one round. As a result, the total transmission delay is $\left \lceil (N-1)/2 \right \rceil \times 2$ time unit.
	\subsubsection{Sun et al.'s MQKA protocol 2 \cite{Sun2015a}}
		Every participant of Sun et al.'s MQKA protocol 2 has to send their qubit sequence to following participant at a round. 2 bits key is agreed when the last following participant sent the sequence back to the participant who prepared the sequence. As a result, the total transmission delay is $N$ time unit.
	\subsubsection{Our proposed protocol}
		The calculation of transmission delay at Our protocol is very simple. It only takes 2 round for 1 bit key agreement. Under the requirement of 2 bit key agreement, the total transmission delay is $2\times 2$ time unit.
	
	\begin{figure}[!t]
		\centering
		\includegraphics[scale=0.5]{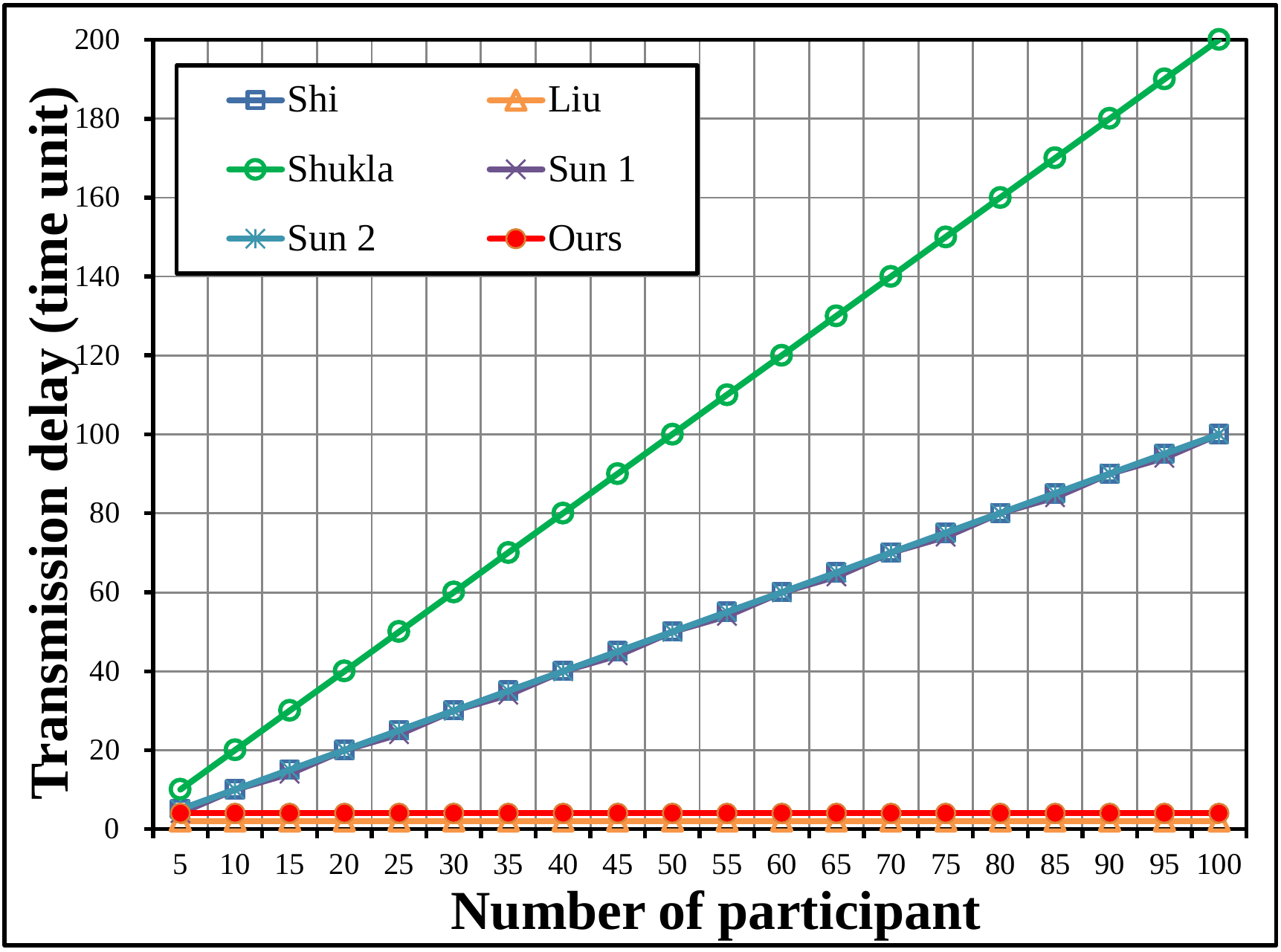}
		\caption{Comparison of transmission delay}
		\label{fig:trans_delay}
	\end{figure}
	Each round should take a time unit. So, even the protocols \cite{Shukla2014,Sun2015,Sun2015a} are great at ``number of qubit measurement". The property of round by round takes more time unit than our protocol. One of them \cite{Liu2012} of ``number of qubit measurement" is less than ours, because each participant can send their self key to others in one round for 1 bit key agreement. But our protocol should takes two. However, our protocol is more efficient than Liu et al. \cite{Liu2012} at other indexes.
	
\section{Conclusion}\label{sec:conclusion}
	This study not only defines the difference of condition \textbf{1} of nowadays QKA protocols, but also proposes a multi-party QKA protocol with multicast method. The consumption comparison section shows the performance that is better than the other protocols such as \cite{Shi2012,Liu2012,Shukla2014,Sun2015a,Sun2015} with unicast method in 4 comparison indexes which are number of ``transmission", ``qubit measurement", ``qubit for channel checking", and ``transmission delay". In addition, the security analysis section shows that this protocol can detect and against the eavesdropper at external and internal attack. As a result, this protocol is the best MQKA protocol, it can not only reduce number of transmission and qubit for channel checking, but also exchange all self keys simultaneously.


%







\bibliographystyle{IEEEtran}
\bibliography{mqka}

\end{document}